\documentstyle{article} \begin{document} \title{
Reply to Comment on "Theory of
Spinodal Decomposition"}\author{S. B.
Goryachev\\ Laboratoire de M\'{e}tallurgie
Physique, B\^{a}t. C6,\\
Universit\'{e} de Lille I,\\ 59655 Villeneuve d'Ascq,
France}
\date{Submitted in Physical Review Letters 06 July 1995\\Revised 18
April
1996} \maketitle

In his Comment \cite{1} to my paper \cite{2}
Rutenberg notes that "the basis of
my analysis of conserved scalar
phase-ordering dynamics, to apply only the global
conservation constraint
\begin{equation} \int {\rm d}{\bf x} \psi=N,
\end{equation} is incorrect".
Because "for physical conserved systems, which
evolve by mass transport,
the stronger local conservation law embodied by
continuity equation" \begin
{equation} \partial \psi/\partial t=- { \bf \nabla j
}. \end {equation}
This question needs clarification.

Replying shortly, I say that my theory
has no any contradiction with the local
conservation law (2).  It describes
the evolution of {\it all three types} of
thermodynamic systems: without
conservation law (NCOP), with global conservation
law (1) (GCOP) and with
local one (2) (LCOP). In these all cases the order
parameter (OP) $\delta
\psi({\bf x},t)=\psi({\bf x},t)-\psi_0({\bf x})$ dynamics
near the extremal
$\psi_0({\bf x})$  is described by the {\it same} evolution
equation
\begin{equation} \partial \psi / \partial t=- \Gamma \delta {\cal F}
/
\delta \psi \end{equation} with {\it different} thermodynamic
potential
functionals ${\cal F} \{ \psi \}$. For NCOP-system ${\cal F} \{
\psi \}\equiv F\{
\psi \}=\int {\rm d}{\bf x} f\{ \psi\} $, where $F\{ \psi
\}$is a free energy
functional and $f\{ \psi \}$ is a specific free energy
one. For GCOP- and
LCOP-systems ${\cal F} \{ \psi \} \equiv \Omega\{ \psi
\}=F-\mu_g N$, where the
{\it constant} $\mu_g$ is a global chemical
potential defined by (1) and
equilibrium condition $\delta {\cal F} /
\delta \psi|_{\psi=\psi_0({\bf x})}=0$.
The difference between $F$ and
$\Omega$ is that the conservation law (1) is taken
into account. It should
be noted that for GCOP and LCOP-systems (1) must be also
used directly for
selection of valid solutions of (3). So {\it there is no
difference between
GCOP and LCOP evolution}. That is why I have used the same
notation (COP)
for them in \cite{2}.

Writing evolution equation (3) I proceed from
classical point of view that a
system has a thermodynamic potential ${\cal
F} \{ \psi \}$ {\it if and only if}
it is related to dynamical class of, so
called, {\it potential} systems. For
potential systems the OP evolution
near the extremal $\psi_0({\bf x})$ is
described by the equation (3). More
generally, ${\cal F} \{ \psi \}$ is the
system's Lyapunov functional for
attractor $\psi_0$, if $\dot{{\cal F}} < 0$ for
$\psi \neq \psi_0$ and
$\dot{{\cal F}} = 0$ for $\psi = \psi_0$. Dynamics of the
potential system
consists of relaxation toward the minimum ${\cal F}_0= {\cal F}
\{ \psi_0
\}$ and the Lyapunov functional existence guarantees a system's
global
asymptotic stability at $\psi = \psi_0$. In more physical language
the condition
$\dot{{\cal F}} < 0$ for $\psi \neq \psi_0$ and $\dot{{\cal
F}} = 0$ for $\psi =
\psi_0$ is nothing more nor less than the second law
of thermodynamics.

Rutenberg states that (2) applies a supplementary
constraint on the OP evolution.
It is not right, because for LCOP we can
always find from (2) and (3) the OP flux
{\bf j} , being a priori an {\it
unknown} quantity, by writing a {\it transport
equation }\cite{3,4}
\begin{equation} {\bf \nabla j}=\Gamma \delta \Omega/ \delta
\psi.
\end{equation} If ${\rm rot}{\bf j}=0$, we can rewrite (4) as
\cite{4}
\begin{equation} {\bf j} =-\Gamma \nabla {\cal M}/ (4 \pi).
\end{equation} Here
${\cal M}({\bf x},t)=\int {\rm d} {\bf x'} \{[\delta
\Omega\{ \psi ({\bf
x'},t)\}/ \delta \psi] / | {\bf x}- {\bf x'}| \}$ is a
non-local characteristic
of the system, so called, scalar potential of the
vortex-free vector field ${\bf
j}({\bf x},t)$ \cite{5}.

In linear
approximation, when $f=[ A \psi^2+C ({\bf \nabla} \psi )^2]/2$, we get
from
(4) and (5) \begin{equation} {\bf \nabla j} = \Gamma (A \psi- C
\triangle
\psi), \end{equation} and $ {\bf j} ({\bf k}, \omega)=\omega_\psi
({\bf k})
\delta \psi ({\bf k}, \omega){\bf k}/k^2 $ with $\omega_\psi
({\bf k})=-{\rm i}
\Gamma (A + C k^2)$. The condition ${\bf j} ({\bf k},
\omega)|_{{\bf k}=0} =0$ is
guaranteed by $\delta \psi ({\bf k},
\omega)|_{{\bf k}=0} =0$ , which follows
from (1).  We see that {\bf j}
consist of { \it two transport modes}: a {\it
dilatation} one, which
corresponds to an extension or a contraction of OP and a
{\it diffusion}
one, which corresponds to Ficken diffusion. If we suppose that
the
characteristic time of OP dilatation $\tau_A=1/(\Gamma A)$ is much
greater
than the caracteristic time of OP diffusion $\tau_C=1/(\Gamma
Ck^2)$, i.e. $ k^2
>> 1 / \xi^2 $ ($\xi= \sqrt{C/A}$ is the correlation
length of OP fluctuations),
we get Fick's law in its classical form
\begin{equation} {\bf j} =-{\cal D} {\bf
\nabla} \psi \end{equation} with
the diffusion coefficient ${\cal D}=\Gamma C$.

To derive Fick's law in the
form \begin{equation} {\bf j} =-\lambda {\bf \nabla}
\mu_l, \end{equation}
where $\mu_l=\mu_l({\bf x},t)$ is a local chemical
potential and $\lambda$
is a transport coefficient, it is necessary to introduce
a {\it local
equilibrium assumption}. Let us divide the system into cells of a
small
volume $V_l=l^d$ with $l<<1/k$. In each cell the diffusion mode of ${\bf
j}
({\bf k}, \omega)$ dies away after the time $\tau_l =l^2 /(\Gamma C)$
much
smaller than the relaxation time $\tau_C$ of the diffusion mode in the
system as
a whole. Then we can believe that all local thermodynamic
variables and functions
as OP $\psi$, free energy $F_l$, entropy $S_l$,
pressure $P_l$, chemical
potential $\mu_l$ etc. are {\it constants} in the
cell. It is the assumption of
local equilibrium that makes possible to
meaningfully define the local free
energy $F_l({\bf x},t)$ that is the same
function of the local thermodynamic
variables as the equilibrium free
energy is of the equilibrium thermodynamic
parameters. That is why the
fundamental differential relation ${\rm d}F_l=-S_l
{\rm d}T-P_l {\rm
d}V+V_l \mu_l {\rm d}\psi$ is valid locally. Replacing $F\{
\psi
\}=\int_{V_l} {\rm d}{\bf x}[ A \psi^2+C ({\bf \nabla} \psi)^2]/2$ by
the
local free energy $F_l(\psi)=V_l A \psi^2 /2$ and using formulas of
traditional
equilibrium thermodynamics we can calculate the local chemical
potential $\mu_l
({\bf x},t) ={\rm d}F_l(\psi)/{\rm d}\psi=V_l A \psi({\bf
x},t)$ and find (8)
with $\lambda=\Gamma C / A$ \cite{4}. If we are
interested by self-organisation
effects, we must consider the situation,
where $\xi^2 k^2 \approx 1$ and ${\bf
j}$ is defined by (6), but not by (7)
or by (8).

We see that Fick's law in form (8) is itself a {\it consequence
} of OP evolution
equation, so it can not be utilised for its {\it
derivation}. Fick's law in form
(7) describes a special case of OP
transport: diffusion without dilatation.
Fick's law in form (8) is an
approximative expression for diffusion flux, not
only in the sense that it
does not take into account the higher order gradient
terms of a chemical
potential $\mu_l$, but also in the sense that it is a law of
the linear
thermodynamics for which the local equilibrium assumption is
compulsory.
General expression for OP flux (5) has nothing to do with the  Fick's
law
form \begin{equation} {\bf j} =-\lambda \nabla \delta F/ \delta
\psi
\end{equation} criticized in \cite{2} and accepted in \cite{1} as
"motivated
phenomenologically". The use of (8) with formally defined "local
chemical
potential" $\mu ({\bf x},t) =\delta F / \delta \psi$ has no
thermodynamic
foundation and is incorrect.  This completely concerns the
equation
\begin{equation} \partial \psi/\partial t=- \lambda {\bf \nabla}^2
\delta F/
\delta \psi.  \end{equation} An origin of the misunderstanding
is, as has been
note in \cite{2}, a false adoption of the Fick's law in
form (9) as a fundamental
general law, which can be used as a basis for
derivation of the OP evolution
equation of nonequilibrium thermodynamic
systems with COP.

Fick's law (8) fails completely in a non-linear
dynamics. In a non-linear
approximation, when $f\{ \psi \}=[ A \psi^2+B
\psi^4/2+C ({\bf \nabla} \psi
)^2+D(\triangle \psi)^2+E(\psi{\bf \nabla}
\psi)^2]/2 $ equation (4) has the form
\begin{equation} {\bf \nabla
j}=\Gamma [ A \psi + B \psi^3-C\triangle \psi
+D\triangle^2 \psi -E\psi^2
\triangle \psi]. \end{equation} We see that ${\bf
\nabla j}$ includes now
not only the dilatation mode and the diffusion mode, but
also a cross term
$E\psi^2 \triangle \psi$, being responsible for the dilatation
- diffusion
coupling. In this case we can not in principal separate dilatation
and
diffusion effects.

Some words about the example given by Rutenberg.  It
does not show any
inconsistence of my theory with the local consevation law
(2), but brilliantly
demonstrate incorrectness of (10).  Let us, following
\cite{1}, take "a special
initial condition" and "require that the
dissipative dynamics be invariant under
$\psi \rightarrow -\psi$ and that
$F\{ \psi \}$ is minimized by $\psi =\pm 1$
everywhere except for a small
sphere where $\psi=-1$, the other of which has
$\psi=+1$ and $ -1$,
respectively". For simplicity, let us omit the gradient term
in Rutenberg's
free energy funtional and take $F\{ \psi \}=\int {\rm
d}{\bf
x}(\psi^2-1)^2$.  Then the system initial state is evidently an
equilibrium one
for all three cases NCOP, GCOP or LCOP. Therefore we must
not observe any changes
of $\psi({\bf x},t)$ in time.  However, following
\cite{1}, we get from (10) that
"the spheres evolve". This is nonsense.
This circumstance can be directly
discovered from (10), if we note that by
the condition ${\bf \nabla}^2 \delta F/
\delta \psi =0$ it defines a false
equilibrium state, being different, in general
case, from the real
equilibrium one.  The later must always minimize $F\{ \psi
\}$, i.e. must
be determined by the condition $\delta F / \delta
\psi=0$.

\end{document}